\documentclass{iaus}
\usepackage{graphics}

\title[Turbulence-Star Formation Connection] 
{MHD turbulence-Star Formation Connection: from pc to kpc scales}

\author[de Gouveia dal Pino \etal\ ]   
{E. M. de Gouveia Dal Pino$^1$,%
\break R. Santos-Lima$^1$,
A. Lazarian$^2$,
M. R. M. Le\~ao$^1$,
\break D. Falceta-Gon\c calves$^{3}$,
 \and G. Kowal$^1$
}

\affiliation{$^1$ IAG, Universidade de S\~ao Paulo, Rua do Mat\~ao 1226,  S\~ao Paulo 05508-090, Brazil
\\ email: {\tt dalpino@astro.iag.usp.br} \\[\affilskip]
$^2$ Astronomy Department, University of Wisconsin, Madison, WI, USA\\[\affilskip]
$^3$ NAC, Universidade Cruzeiro do Sul, Rua Galv\~ao Bueno
868,  S\~ao Paulo 01506-000, Brazil
}

\pubyear{2011}
\volume{274}  
\pagerange{}
 \jname{Advances in Plasma Astrophysics}
\editors{A. Bonanno, E. de Gouveia Dal Pino, A. Kosovichev, eds.}
\begin{document}

\maketitle

\begin{abstract}
The transport of magnetic flux to outside of collapsing molecular clouds is a required step to allow the formation of stars. Although ambipolar diffusion is often regarded as a key mechanism for that, it has been recently argued that it may not be efficient enough. In this review, we discuss the role that MHD turbulence plays in the transport of magnetic flux in star forming flows. In particular, based on recent advances in the theory of fast magnetic reconnection in turbulent flows, we will show results of three-dimensional numerical simulations that indicate that the diffusion of magnetic field induced by turbulent reconnection can be a very efficient mechanism, especially in the early stages of cloud collapse and star formation. To conclude, we will also briefly discuss the turbulence-star formation connection and feedback in different astrophysical environments: from galactic to cluster of galaxy scales.

\end{abstract}

\firstsection 
\section{Introduction}

It is generally believed that  stars form within molecular clouds. This is a natural consequence of their low temperatures and high densities which help the gravitational force to overcome the internal pressures that act ``outward'' to prevent collapse.

The  internal motions in the molecular clouds are governed by turbulence in a cold, magnetized gas. Broad line widths ranging from a few to more than 10 times the sound speed are observed in the molecular clouds and the  interstellar magnetic field strength is a few microgauss, in rough equipartition with the kinetic energy in the interstellar medium. For this reason the turbulent motions are regarded as  mainly supersonic and trans-Alfv\'enic (\cite[Elmegreen \& Scalo 2004]{ElmegreenScalo2004}; \cite[Heiles \& Troland 2005]{HeilesTroland2005}). The  role of this turbulence in the interstellar dynamics and star formation, although amply discussed in the literature (see, e.g., the reviews by \cite[Elmegreen \& Scalo 2004]{ElmegreenScalo2004}; \cite[McKee \& Ostriker 2007]{McKeeOstriker2007}), is still poorly understood.

A vital question that frequently permeates these debates is the diffusion of the magnetic field through a collapsing cloud. If the magnetic field were perfectly frozen into the interstellar gas during the gravitational collapse, the magnetic field strength of a typical star like our Sun would be more than 10 orders of magnitude larger than we observe today. Thus, there must be some mechanisms that are effective in removing the excess of magnetic flux during the star formation process. To address this problem, researchers usually appeal to the ambipolar diffusion (AD) of the magnetic field through the neutral component of the plasma (\cite[Mestel \& Spitzer 1956]{MestelSpitzer1956}; \cite[Shu 1983]{Shu1983}; \cite[Tassis \& Mouschovias 2005]{TassisMouschovias2005}). However, it is unclear yet if AD can be high enough to provide  the magnetic flux transport in collapsing fluids (\cite[Shu \etal\ 2006]{Shu_etal2006}; \cite[Krasnopolsky \etal\ 2010]{Krasnopolsky_etal2010}).

Here, we discuss an alternative mechanism  based on turbulent magnetic reconnection that seems to be able to provide an efficient magnetic flux removal, from the initial stages of accumulating interstellar gas to the final stages of the accretion to form protostars.

\section{Magnetic Flux transport through turbulent reconnection}

The idea that, due to turbulence, the magnetic field lines can reconnect and then lead to the removal of magnetic flux from collapsing clouds was first discussed by \cite[Lazarian (2005)]{Lazarian2005} and more recently explored numerically by \cite[Santos-Lima \etal\ (2010)]{Santos-Lima_etal2010}. This was based on the magnetic reconnection model of \cite[Lazarian \& Vishniac (1999)]{LazarianVishniac1999} which was successfully confirmed by three-dimensional numerical simulations in \cite[Kowal \etal\ (2009)]{Kowal_etal2009}. This model establishes  that in the presence of \textit{weak} turbulence, reconnection of the field lines is \textit{fast} since many reconnection events occur simultaneously at smaller and smaller scales (even with  Ohmic resistivity only). As a consequence, a robust removal of magnetic flux can be expected both in partially and fully ionized plasma.

\section{Three-Dimensional simulations of magnetic field diffusion by turbulent reconnection  in gravitating clouds}

\begin{figure}
\centering
\resizebox{4.8cm}{!}{\includegraphics{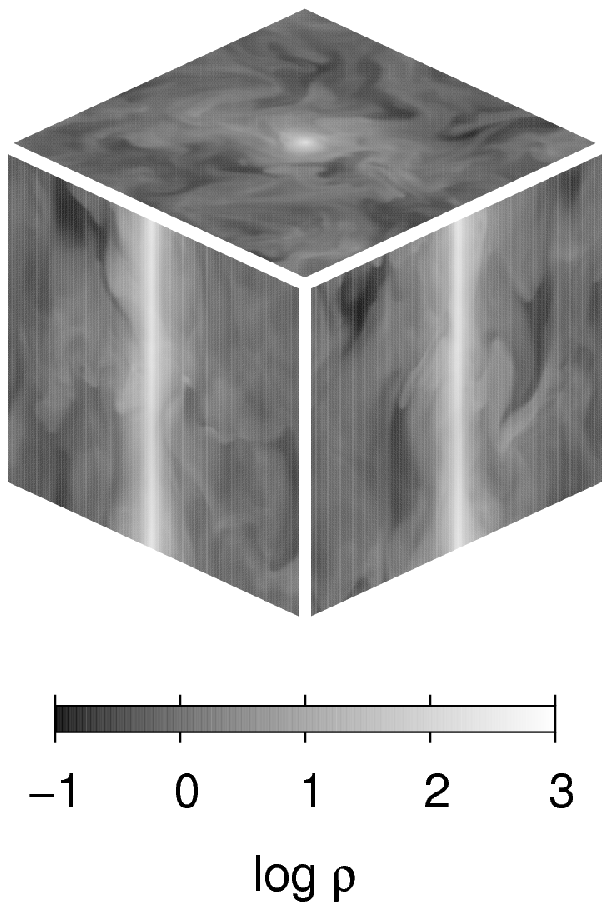} }
\resizebox{7.0cm}{!}{\includegraphics{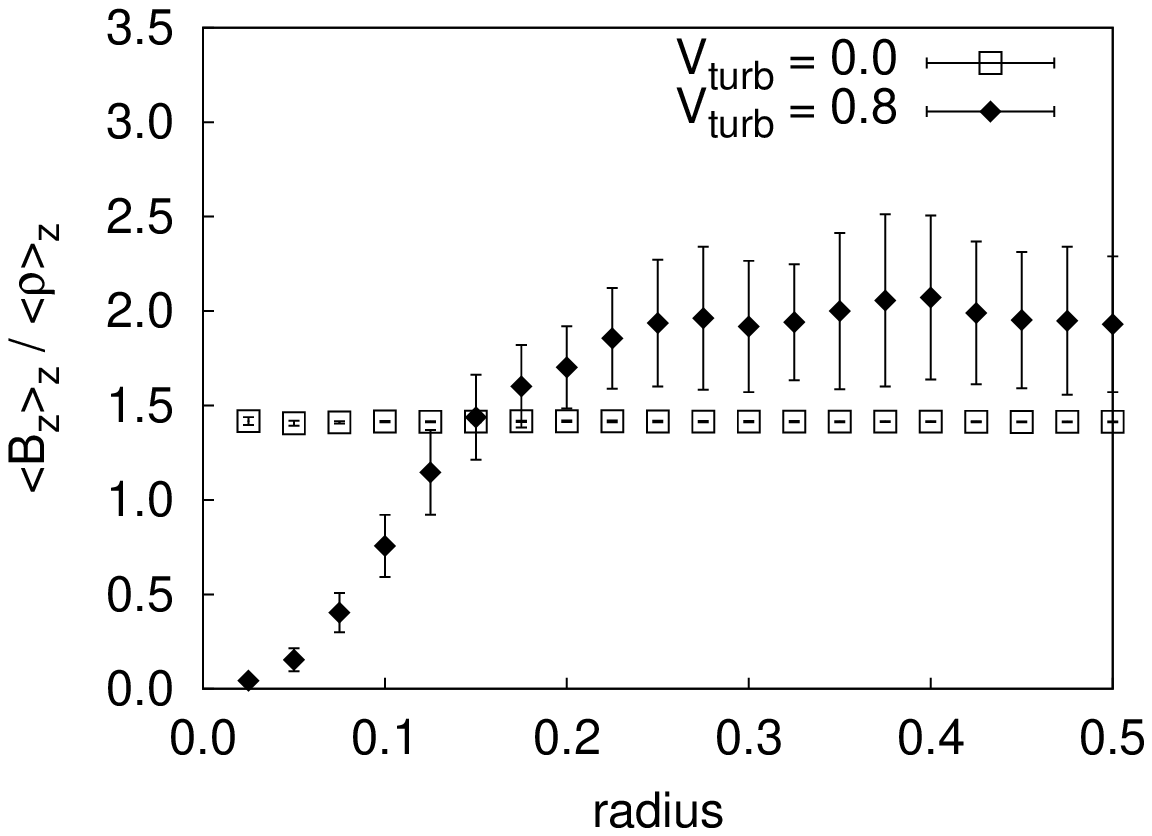} }
\caption[]{Left: Logarithm of the density field in $t=3$ c.u. (where 1 time code unit is given by the box size $L$ divided by $c_s$ (with $c_s=1$ c.u. being the sound speed) for  a system initially in magneto-hydrostatic equilibrium, in a cylindrically symmetric gravitational potential,  with a constant ratio between the thermal and the magnetic pressure ($\beta = 1$),  and $v_A/c_s = 1.4$ c.u., where $v_A$ is the Alfv\'en speed. The  gravitational potential is $\Psi \sim A/R$, where $R$ is the radial distance to the center of the cloud, and $A$ is a gravitational parameter given in units of $c_s^2 L$.  In this case, $A=0.9$ c.u. The central xy, xz, and yz slices of the system were projected on the respective walls of the cubic computational domain.
Right: Ratio between the magnetic field and the density (averaged over the z direction) normalized by the initial value, as a function of the cloud radius for a collapsing system (starting out of the magneto-hydrostatic equilibrium)  both with and without turbulence, in $t=8$ c.u. The cloud has  the same initial conditions as those of the left system. The magnetic field is given in units of $B= c_s \sqrt{4 \pi \rho}$.  Error bars show the standard deviation. The numerical resolution is $256^3$. Extracted from \cite[Santos-Lima \etal\ (2010)]{Santos-Lima_etal2010}.}
\label{fig1}
\end{figure}

We have performed 3-D MHD simulations employing a modified version of the Godunov-based MHD code developed by \cite[Kowal \etal\ (2007)]{Kowal_etal2007}. We first considered \textit{cubic} cloud systems as shown if Figure \ref{fig1}, subject to a gravitational potential with cylindrical geometry, an isothermal equation of state, and immersed in an initial vertical magnetic field. The boundaries were assumed to be periodic and the simulations were started with the cloud either in magneto-hydrostatic equilibrium (Figure \ref{fig1}, left)  or already collapsing under the effect of the gravitational potential.  In all cases,  transonic, sub-Alf\'enic, non-helical turbulence with an rms velocity ($V_{rms}$) around unity was injected in the system which was then let to evolve (see \cite[Santos-Lima \etal\ 2010]{Santos-Lima_etal2010} for details).

As an example, Figure \ref{fig1} (right) shows the  ratio between the magnetic field and the density (both averaged over the z direction) as a function of the cloud radius for a collapsing system  with and without turbulence. For a cylindrical symmetry, this ratio gives a measure of the magnetic flux to mass ratio of the system. The results clearly show that in the presence of weak turbulence there is an efficient removal of magnetic flux from the central regions  (where gravity is stronger and the density is higher) to the outer regions of the system. The fact that in the absence of turbulence there is no change in the magnetic-flux-to-mass ratio is a clear evidence that in the case with turbulence, the transport is due to turbulent reconnection diffusion. The same effect was detected for clouds starting in magneto-hydrostatic equilibrium. This is an indication that the process of turbulent magnetic field removal should be applicable both to quasi-static subcritical molecular clouds and to collapsing cores. These results were also found to be insensitive to the numerical resolution. Tests made with resolutions of $128^3$, and $512^3$ gave essentially the same results as those of $256^3$ (shown in the figures) thus confirming the robustness of the results above.

We have also examined an extensive parameter space considering different strengths of the gravitational potential, the turbulent velocity and the magnetic field. The increase of the gravitational potential as well as the magnetization of the gas (decrease of $\beta$) increases the efficiency of the transport of magnetic flux to the periphery of the system. This is expected, as turbulence brings the system to a state of minimal energy. The effect of varying magnetization in some sense is analogous to the effect of varying gravity. The physics is simple, the lighter fluid (magnetic field) is segregated from the heavier fluid (gas), supporting the notion that the reconnection-enabled diffusivity relaxes the magnetic field + gas system in the gravitational field to its minimal energy state.

When the turbulent velocity is increased, there is at first a trend to remove more magnetic flux from the center allowing a more efficient infall of matter to the center. However, if we keep increasing the turbulent velocity, there is a threshold above which cloud collapse fails. Gravity becomes negligible compared with the turbulent kinetic energy and matter is removed from the center along with the magnetic flux and this causes cloud fragmentation rather than its collapse to form stars. Indeed, this effect may be very frequent in the ISM of our Galaxy since the star formation efficiency is known to be very small.

An enhanced Ohmic resistivity of unknown origin has been invoked in the literature as a way to remove magnetic flux from cores and accretion disks (e.g., by \cite[Shu \etal\ 2006]{Shu_etal2006}). We have then also performed numerical simulations of systems without turbulence but with  enhanced Ohmic diffusivity. The comparison of these models with those with turbulent have shown that turbulent reconnection diffusivity can mimic the effects of an enhanced Ohmic resistivity in gravitating clouds.

\begin{figure}
\centering
\resizebox{6.5cm}{!}{\includegraphics{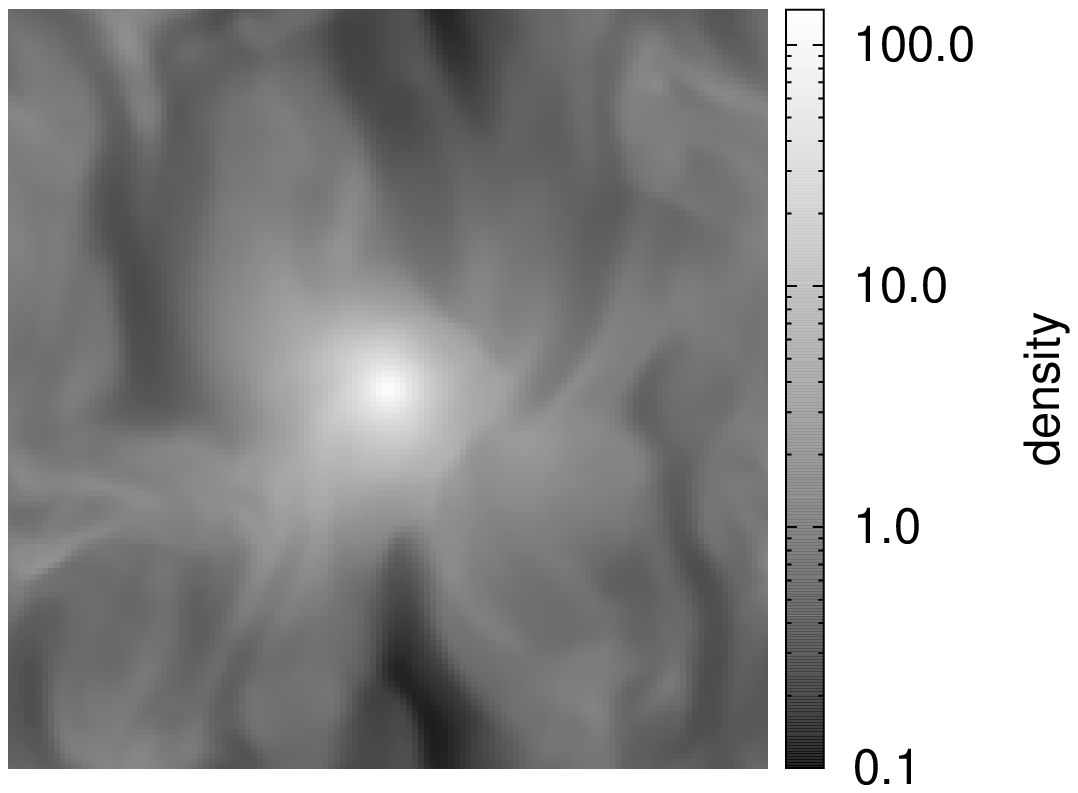} }
\resizebox{6.5cm}{!}{\includegraphics{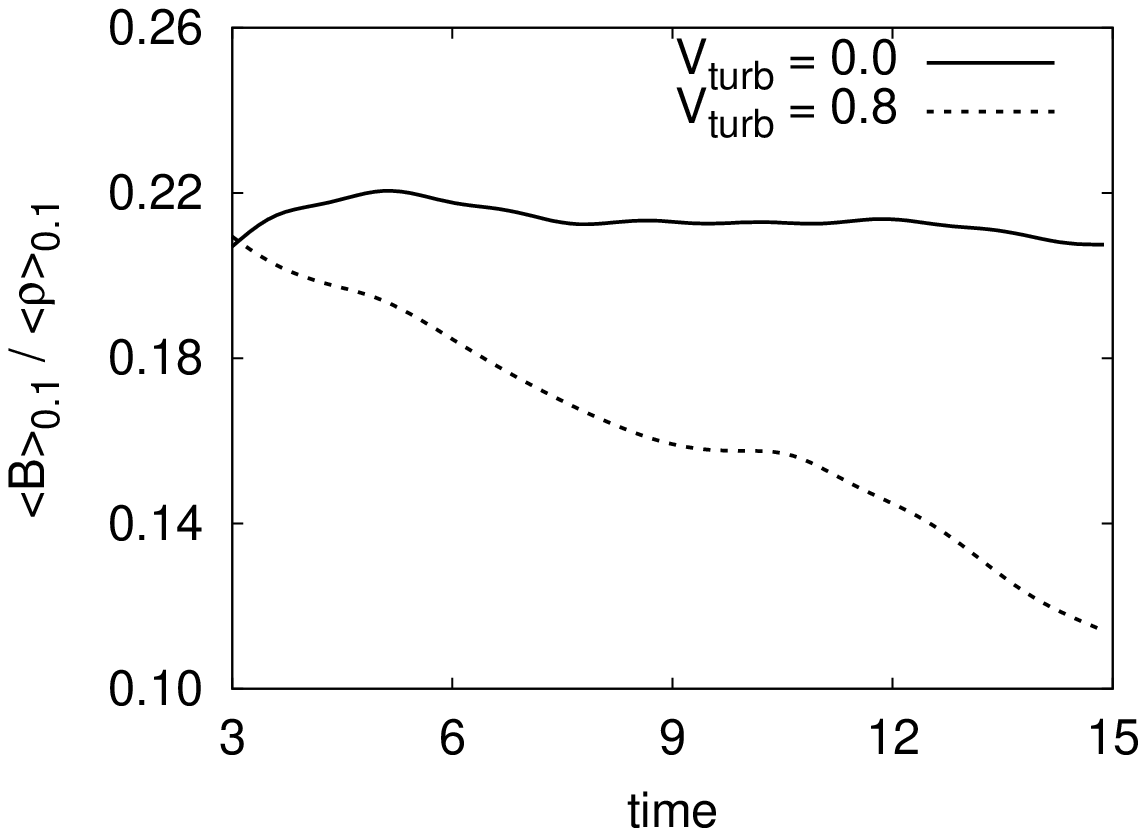} }
\caption[]{Left panel: density map of the central slice of a collapsing cloud with a spherically symmetric potential (central mass M $= 50$ M$_{\odot}$) in  $t=10$ c.u. $= 100$ Myr. The initial magnetic field is in the vertical direction of the image and has an intensity of $0.1 \mu$G. The initial density in the medium is $1$ c.u. $=100$ cm$^{-3}$, and $\beta=3$. Right panel: time evolution of the magnetic field-to-density ratio, averaged within the core region of radius $R=0.1$ c.u. = $0.3$ pc. The curves were smoothed for making the visualization clearer. (\cite[Le\~ao \etal\ 2011, in prep.]{Leao_etal2011})}
\label{fig3}
\end{figure}

Recently, we have been testing more realistic cloud systems, considering  spherically symmetric potentials. The effects of self-gravity, which become more significant at the late stages of the collapse, have been also introduced. These assumptions are consistent with observed evidence that the large, star-forming clouds are confined  by their own gravity (like stars, planets, and galaxies) rather than by external pressure (like clouds in the sky). This evidence comes from the fact that the ``turbulent'' velocities inferred from CO line-widths scale in the same manner as the orbital velocity (like in virial systems). Figure \ref{fig3} shows a snapshot of one of these preliminary tests and the temporal evolution of the magnetic field-to-density ratio averaged within a region of radius $R=0.1$ c.u., where $L=1$ c.u. $= 3$ pc is the size of the box. Again, it indicates an efficient removal of magnetic flux in comparison to the case when no turbulence is injected.

\section{The role of turbulent reconnection on the formation of rotationally supported circumstellar disks}

In the late stages of star formation, former studies showed that the observed  magnetic fields in molecular cloud cores (which imply magnetic flux-to-mass ratios of the order of a few times unity; \cite[Crutcher 2005]{Crutcher2005}) are high enough to inhibit the formation of rationally supported disks during the main protostellar accretion phase of low mass stars (see e.g., \cite[Krasnopolsky, Li, and Shang 2010]{KrasnopolskyLiShang2010} and references therein).

For realistic levels of core magnetization and ionization, recent work has demonstrated that AD is not sufficient to weaken the magnetic braking. Further, \cite[Krasnopolsky, Li, and Shang (2010)]{KrasnopolskyLiShang2010} have found that in order to enable the formation of rotationally supported disks during the protostellar mass accretion phase, a resistivity a few orders of magnitude larger than the classic microscopic resistivity values ($\eta \sim 10^{19}$ cm$^2$s$^{-1}$) would be needed.

\begin{figure}
\centering
\resizebox{7.0cm}{!}{\includegraphics{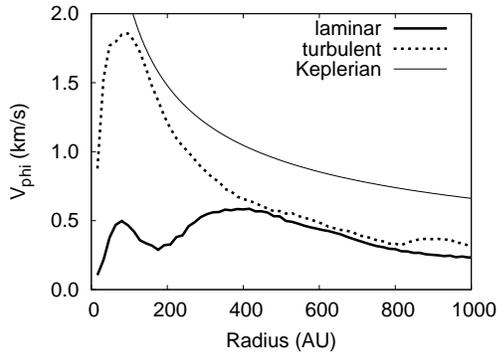} }
\caption[]{Radial profile of the rotational velocity of a protostellar disk  with and without turbulence in $t = 1.5 \times 10^{12}$ s. The Keplerian profile is shown for comparison. The progenitor cloud has initially  uniform density $\rho = 1.4 \times 10^{-19}$ g cm$^{-3}$ and  uniform magnetic field parallel to the rotation axis with intensity $B = 100 \mu $G. The sound speed is $c_s = 2 \times 10^{4}$ cm s$^{-1}$. The central proto-star has a mass of $0.5$ M$_{\odot}$. Extracted from \cite[Santos-Lima, de Gouveia Dal Pino, Lazarian 2011 (in prep.)]{Santos-LimadeGouveiadalPinoLazarian2011} .}
\label{fig5}
\end{figure}

We have also explored the effects of turbulent reconnection diffusion on removing magnetic flux during the formation of these disks. We found that turbulent magnetic reconnection diffusivity causes an efficient transport of magnetic flux to the outskirsts of the disk at time scales compatible with the accretion time scales, allowing the formation of a rotationally supported protostelar disk with nearly Keplerian profile (see Figure \ref{fig5}, from \cite[Santos-Lima, de Gouveia Dal Pino, Lazarian 2011, in prep.]{Santos-LimadeGouveiadalPinoLazarian2011}).

\section{Star formation turbulence feedback}

So far, we have focused on the effects of turbulence on star formation (SF). Now, to conclude, we would like to address very briefly the SF feedback on the ISM and environment at galactic scales. This  is a fundamental issue in the evolution of galaxies. SF in galaxies depends on both the gas content and the energy budget of the ISM. Since the most efficient stellar energy power is exerted by supernovae (SNe) and, particularly by the explosions of shortly living massive stars as type II SNe, their feedback is of fundamental relevance. There has been extensive investigation about the regulation of the ISM by the SF feedback through SN bubbles (see \cite[de Avillez 2000]{deAvillez2000}; \cite[de Avilez \& Breitschwerdt 2005]{deAvilezBreitschwerdt2005}; \cite[Melioli \& de Gouveia Dal Pino 2004]{MeliolideGouveiaDalPino2004}; \cite[Melioli \etal\ 2005]{Melioli_etal2005}; \cite[Hensler 2010]{Hensler2010} and references therein).

Likewise, galactic winds and outflows emerging from star forming galaxies are also believed to be driven mainly by SN turbulence (\cite[Strickland \& Stevens 2000]{StricklandStevens2000}; \cite[Tenorio-Tagle \etal\ 2003]{Tenorio-Tagle_etal2003}; \cite[Cooper, Bicknell \& Sutherland 2008]{CooperBicknellSutherland2008}; \cite[Melioli \etal\ 2008]{Melioli_etal2008}, \cite[2009]{Melioli_etal2009}).

Recently, we have examined the role of SF/SN driven turbulence also at the scales of galaxy cluster cores. For instance, Perseus, the brightest galaxy  cluster observed in X-ray, presents a rich structure of cold filaments  and loops seen in stellar continuum, H$\alpha$,  and [NII] line emission, which are distributed all around the central galaxy of the cluster (NGC1275) extending for up to 50 kpc (\cite[Fabian \etal\ 2008]{Fabian_etal2008}). The origin of the filaments is still unknown. We performed 2.5 and 3-dimensional MHD simulations of the central region of the cluster in which turbulent energy, triggered by star formation and supernovae (SNe) explosions was introduced. The simulations have revealed that the turbulence injected by massive stars could be responsible for the nearly isotropic distribution of filaments and loops that drag magnetic fields upward as indicated by the observations. Weak shell-like shock fronts propagating into the intra-cluster medium with velocities of $100-500$ km/s were found, also resembling the observations.

As the turbulence is subsonic over most of the simulated volume, the turbulent kinetic energy is not efficiently converted into heat and additional heating is required to suppress the cooling flow at the core of the cluster.
NGC 1275, the central galaxy in the Perseus cluster, is the host of gigantic
hot bipolar bubbles inflated by the AGN jets observed in the radio and it is generally believed  that this is the main agent to stop the cooling flow (see also Feretti et al. in this volume).   Simulations combining the MHD turbulence with the AGN outflow were able to reproduce the temperature radial profile observed around NGC1275. While the AGN is the main heating source, the supernova turbulence seems to be an important mechanism  to isotropize the energy distribution  besides being able to explain the rich filamentary and weak shock structures (\cite[Falceta-Gon\c calves \etal\ 2010a]{Falceta-Goncalves_etal2010a}, \cite[2010b]{Falceta-Goncalves_etal2010b}).

\section{Summary}

We  first addressed the effects of turbulence on star formation (SF) and discussed a new possible mechanism for magnetic field flux removal from collapsing clouds and cores based on  MHD turbulent reconnection, which seems to be more efficient than ambipolar diffusion. 3D numerical modeling sweeping an extensive parameter space has revealed that turbulent reconnection can play an important role in the removal of  magnetic field flux in different stages of SF and make molecular clouds which are initially subcritical to become supercritical and therefore, ready to collapse. Besides, this mechanism can also  help the transport of magnetic flux to the outer regions of protostellar disks  allowing the formation of rotationally supported disks.

We have also briefly addressed the star formation and supernovae-driven turbulence feedback into the environment at galactic scales. 3D MHD simulations revealed that this can be a mechanism very efficient to generate outflows, and cold, magnetized filaments and clouds even in the cores of galaxy clusters.

\end{document}